\documentclass{elsart}

\usepackage{epsfig}

\usepackage{amssymb}

\begin{document}
\begin{frontmatter}

\title{On the interpretation of muon-spin-rotation experiments in the mixed state of type-II superconductors}

\author[label1,label2]{I. L. Landau}
\author[label1]{H. Keller}
 \address[label1]{Physik-Institut der Universit\"at Z\"urich, Winterthurerstrasse 190, CH-8057 Z\"urich, Switzerland}
 \address[label2]{Institute for Physical Problems, 117334 Moscow, Russia}



\begin{abstract}
	
We argue that claims about magnetic field dependence of the magnetic field penetration depth $\lambda$, which were made on the basis of muon-spin-rotation ($\mu$SR) studies of some superconductors, originate from insufficient accuracy of theoretical models employed for the data analysis. We also reanalyze some of already published experimental data and demonstrate that numerical calculations of Brandt [E.H. Brandt, Phys. Rev. B {\bf 68}, 54506 (2003)] may serve as a reliable and powerful tool for the analysis of $\mu$SR data collected in experiments with conventional superconductors. Furthermore, one can use this approach in order to distinguish between conventional and unconventional superconductors. It is unfortunate that these calculations have practically never been employed for the analysis of $\mu$SR data.

\end{abstract}

\begin{keyword}

type-II superconductors \sep mixed state \sep unconventional superconductivity  \sep muon-spin-rotation experiments  \sep magnetic field penetration depth

\PACS 74.70.Dl \sep 74.25.Op, 74.25.Ha,76.75.+i

\end{keyword}
\end{frontmatter}


\section{Introduction}

Muon-spin-rotation ($\mu$SR) experiments in the mixed state of type-II superconductors provide unique information about superconducting properties of the investigated sample. An important advantage of this method is that muons probe the bulk of the sample and therefore, the results are not distorted by possible imperfections of the sample surface.  At the same time, in order to extract quantitative results from  $\mu$SR measurements, a detailed model of the magnetic field distribution in the mixed state is needed. As well as we are aware, only the Ginzburg-Landau (GL) theory \cite{GL} of the Abrikosov vortex lattice \cite{abr} is developed to such a level \cite{brandt88,brandt97,brandt03}. As was recently demonstrated, if an adequate model is available, not only the magnetic field penetration depth $\lambda$ but also the upper critical field $H_{c2}$ can be found from $\mu$SR data  collected in different applied magnetic fields \cite{khas-l}. It has to be remembered, however, that theoretical calculations of Refs. \cite{brandt88,brandt97,brandt03} are related to superconductors with one and isotropic energy gap only.   This is why, this kind of analysis should be used with extreme caution in the case of unconventional superconductors, in which the applicability of theoretical models is not obvious. 

We also point out a very interesting and promising approach which was developed in Refs. \cite{ishi1,ishi2,ishi3,ishi4}. In these works a microscopic theory was used for calculation of the mixed state parameters. An important advantage of this approach is that the results are not limited to conventional superconductors and it can be used at temperatures well below $T_c$ both for $s$- and $d$-wave pairing. 

In recent years, $\mu$SR measurements were widely used for studying of different unconventional superconductors such as  high-$T_c$ materials, MgB$_2$ and others. Some very interesting results were obtained. It was demonstrated that in some cases the magnetic field penetration depth $\lambda$ and the superconducting coherence length $\xi$, evaluated from $\mu$SR measurements, depend on the applied magnetic field (see, e.g., \cite{son97,son97-3,son99,ohi,son03,son97-2,son04,kado,kadorev}). This result, however, contradicts the GL theory, which was used as a basis for the data analysis. This contradiction is a clear sign that the corresponding models are not adequate for describing the magnetic field distribution in the mixed state of these compounds and rises the question about physical meanings of $\lambda(H)$ and $\xi(H)$ obtained in such a way. As we argue below, magnetic field dependences of $\lambda$ and $\xi$ cannot be obtained from $\mu$SR experiments if the conventional GL theory or the London model were employed for the analysis of experimental data. Moreover, because in the mixed state the superconducting order parameter is not spatially uniform, there is no reasonable way to define either $\lambda$ or $\xi$. In other words, the physical meanings of magnetic field dependences of $\lambda$ and $\xi$, evaluated from $\mu$SR data, are quite different from traditional definitions of these two lengths. This circumstance was recognized in Refs. \cite{sonson,son05,son06} where it was pointed out that $\lambda(H)$, evaluated in such a way, represents some fit-parameter rather than the magnetic field penetration depth. We underline that the same should also be addressed to $\xi(H)$ dependences. In the following section, in order to avoid confusion, we shall use $\lambda_0$ and $\xi_0$ to denote values $\lambda$ and $\xi$ for $H \rightarrow 0$.

\section{Conventional superconductors}

Superconductors with $s$-pairing and one energy gap we shall consider as conventional, independent of their pairing mechanism. Because the GL theory is traditionally used for analyses of $\mu$SR data, we limit our consideration to this theory.

The magnetic field penetration depth $\lambda_0$ together with the zero-field coherence length $\xi_0$ represent two fundamental lengths of the GL theory. If their values for some particular temperature $T$ are known, one can calculate the GL parameter 
\begin{equation}
\kappa(T) = \lambda_0(T)/\xi_0(T),
\end{equation}
the thermodynamic critical magnetic field
\begin{equation}
H_c(T)= \frac{\Phi_0}{2\sqrt{2}\pi\lambda_0(T)\xi_0(T)},
\end{equation}
the upper critical field
\begin{equation}
H_{c2}(T)= \sqrt{2}\kappa H_c(T)= \frac{\Phi_0}{2\pi\xi_0^2(T)},
\end{equation}
the lower critical field
\begin{equation}
H_{c1}(T)= \frac{\ln \kappa(T) + \alpha(\kappa)}{\sqrt{2}\kappa(T)}H_c(T) = [\ln \kappa(T) + \alpha(\kappa)]\frac{\Phi_0}{4\pi\lambda_0^2(T)}
\end{equation}
with $\alpha(\kappa) = 0.49693 + \exp[ - 0.41477 - 0.775\ln \kappa - 0.1303(\ln \kappa)^2]$ \cite{brandt03} Furthermore, in the case of conventional superconductors, any characteristics of the sample for any value of an applied magnetic field may also be calculated and expressed via $\lambda_0$ and $\xi_0$. Very detailed numerical calculations of different parameters of the mixed state for a very wide range of $\kappa$ and for magnetic fields ranging from $H_{c1}$ to $H_{c2}$ are presented in Ref. \cite{brandt03}. 

Muons probe the distribution of the magnetic induction in the sample. In high-$\kappa$ superconductors and low magnetic inductions $B$, contributions of vortex cores can be neglected (London limit) and the distribution of the magnetic induction around a single vortex line may be written as
\begin{equation}
B(r) =\frac{\Phi_0}{2\pi\lambda_0^2} K_0(r/\lambda_0),
\end{equation}
where $r$ is the distance from the vortex center, $\Phi_0$ is the magnetic flux quantum and $K_0$ is the modified Bessel function. Because Eq. (5) is obtained from the London theory, it gives an unphysical divergence of $B$ at $r = 0$. In order to improve Eq. (5), an appropriate cutoff has to be introduced \cite{clem75,hao91,brandt95}. It should be remembered, however, that the results of Refs. \cite{clem75,hao91,brandt95} can be considered as sufficiently accurate in low magnetic fields $H \ll H_{c2}$ only. If this condition is not satisfied, numerical solution of the GL equations must be used for a reliable analysis $\mu$SR data. The magnetic induction distribution may be calculated  as a linear superposition of inductions created by different vortices (see, for instance, Ref. \cite{brandt95}). 

By measuring muon relaxation rates, one obtains the distribution of the magnetic induction $P(B)$ experimentally, which allows to calculate the variance of the magnetic induction
\begin{equation}
\sigma^2 = \left( \overline{B^2(r) - \overline{B^2}}\right),
\end{equation}
where $\overline{\cdots} = (1/V)\int\cdots d^3r$  means spatial averaging over superconductor of volume $V$. If the distribution of the magnetic induction around vortices is known, $\sigma$ can also be calculated theoretically. According to \cite{brandt03}
\begin{equation}
\sigma =  F(\kappa,B/B_{c2})/\lambda_0^2,
\end{equation}
where the parameter $F$ depends on $\kappa$ and $B/B_{c2}$. If the value of $F$ is known, $\lambda_0$ may straightforwardly be evaluated. In the case of $\kappa \gg 1$ and $b \ll 1$, $F \approx 0.061\Phi_0$. In other situations, reliable results can be obtained from Ref. \cite{brandt03}. Eq. (7) may also be written as $\sigma =  (2\pi H_{c2}/\Phi_0) F(\kappa,B/B_{c2})/ \kappa^2$. This representation may be convenient if evaluation of $\kappa$ is preferable. 

While the zero-field value of $\lambda$ enters the theory, the actual magnetic field penetration depth is field dependent. According to the original Ginzburg and Landau publication \cite{GL}, if the magnetic field is parallel to the sample surface,
\begin{equation}
\lambda(H) = \lambda_0\left[1 + f(\kappa)H/H_c\right].
\end{equation}
The function $f(\kappa)$ grows monotonically with $\kappa$ in such a way that  for $\kappa \ll 1$ $f(\kappa) \sim \kappa/4\sqrt{2}$ and $f(\infty) = 0.125$ \cite{GL}. Taking into account Eq. (1), we see that even the magnetic field dependence of $\lambda$ may be expressed via $\lambda_0$ and $\xi_0$. The  $\lambda(H)$ dependence arises due to suppression of the order parameter $|\psi|$ by the applied magnetic field. In bulk type-II superconductors, Eq. (8)  is applicable in the Meissner state only, i.e., in magnetic fields $H < H_{c1}$. If $H \ge H_{c1}$, the magnetic field penetrates into the bulk of the sample forming a lattice of Abrikosov vortices.

\begin{figure}[t]
\begin{center}
  \epsfxsize=0.8\columnwidth \epsfbox {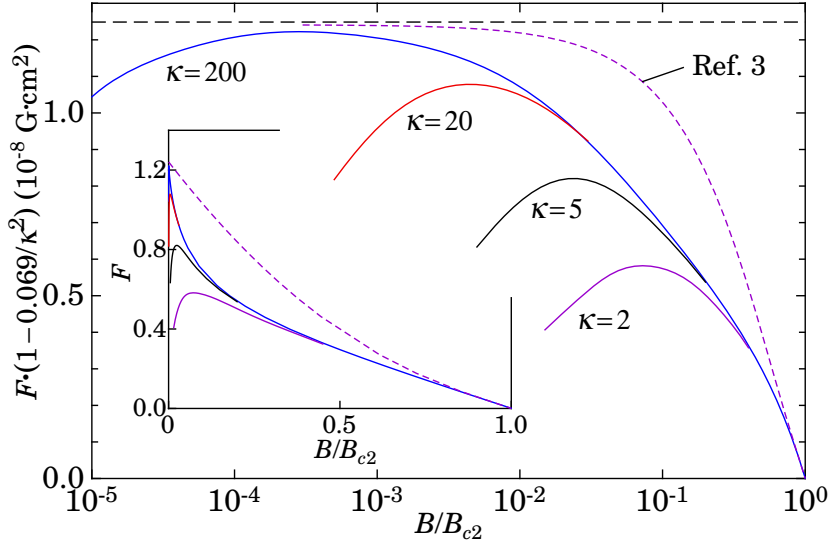}
  \caption{$F$ multipied by $(1-0.069/\kappa^2)$ as a function of $B/B_{c2}$ according to \cite{brandt03}. The dashed line shows the $F(B/B_{c2})$ according to interpolation formula proposed in \cite{brandt88}. The horizontal line corresponds to $F = 0.061\Phi_0$. The inset shows the same curves on linear scales.}
 \end{center}
\end{figure}
If spatial variations of the order parameter can be neglected, the magnetic induction decays exponentially on the flat surface of the sample. In the case of cylindrical geometry (around vortices), the same decay is described by the Bessel function (see Eq. (5)). Considering $\mu$SR experiments in the mixed sate of type-II superconductors, we can use Eq. (5) if the total volume of vortex cores is negligibly  small compared to the volume of the sample, i.e., $\kappa \gg 1$ and $B \ll B_{c2}$. If one or both of these conditions is not satisfied, spatial variations of the order parameter have to be taken into account. Numerical calculations of Brandt are shown in Fig. 1. If $\kappa \gg 1$, the function $F(B/B_{c2})$ is practically independent of $\kappa$. As may be seen in Fig. 1, $F$ remains magnetic field dependent even for very small values of  $B/B_{c2}$. The maximum on the $F(B/B_{c2})$ dependence at $B/B_{c2} \approx 0.17/\kappa^{1.2}$ is the obvious consequence of the fact that $\sigma$ vanishes at $B/B_{c2} \rightarrow 0$. At very low magnetic inductions $F$ is proportional to $\sqrt{(B/B_{c2})}$ \cite{brandt03}.

It must be clearly understood that, although the applied magnetic field influences properties of the sample both in the Meissner and the mixed states, the physics of this influence is completely different. In the Meissner state, supercurrents are induced in the surface layer of the sample. The density of these currents is proportional to $H$ and they depress the order parameter, which leads to an increase of $\lambda$ (see Eq. (8)). Because the reduction of the order parameter $| \psi |$ is small ($|\Delta \psi| \ll |\psi|$), one may still introduce the magnetic field penetration depth in its traditional way. 

In the mixed state, the situation is completely different. Because there are no currents, which are proportional to $H$, the absolute value of the applied magnetic field is irrelevant. Only the distance between vortices given by the magnetic induction $B$ is important. At magnetic inductions $B \lesssim 0.1B_{c2}$, overlapping of vortex cores may be neglected and vortex properties are independent of the applied magnetic field \cite{brandt03}. Only the vortex density is changed. At higher magnetic inductions, vortex cores overlap and not only the vortex density, but also properties of individual vortices are magnetic field dependent. Because the local value of the magnetic field penetration depth is inversely proportional to the modulus of the order parameter $|\psi|$, there is no much sense to introduce any unique value of $\lambda$ corresponding to each particular magnetic field. The correct approach is to calculate some measurable quantities theoretically and compare them with experimental results. In this way, however, only the zero-field value of $\lambda$ can be evaluated. If the value of $\lambda$ resulting from the analysis of experimental data depends on the applied magnetic field, it means that the theory, which was employed for the analysis, does not describe the actual experimental situation and the approach to the analysis should be reconsidered.

If the magnetic field dependence of $F$ is not taken into account or accounted for incorrectly, the analysis of muon depolarization rates would result in some effective $\lambda_{eff}$, which is magnetic field dependent. The knowlege of $\lambda_{eff}(H)$, however, does not represent any particular interest. This is why it is important to use reliable models of the mixed sate in order to evaluate $\lambda_0$. 

As is well known, the GL theory is formally applicable at temperatures close to $T_c$ only. This is why quantitative applicability of theoretical calculations to the analysis of experimental data at  temperatures well below $T_c$ is not obvious. However, as it was recently demonstrated, the magnetic field dependence of $\sigma$ at $T \rightarrow 0$ can be very well fitted by calculations of Brandt with two fit-parameters $\lambda_0$ and $H_{c2}$ \cite{khas-l}. Moreover, the value of $H_{c2}$, evaluated in such a way, coincides with the result of magnetization measurements. We consider this as a proof that the theoretical $\sigma(H)$ dependence calculated in framework of the GL theory can indeed be used for quantitative analysis of isothermal experimental data even at temperatures $T \ll T_c$. Below we reconsider several experimental $\mu$SR studies and demonstrate that their results may perfectly be described by the conventional GL theory although the magnetic field dependence of $\lambda$ was claimed in some of the original publications.

\section{Unconventional superconductors}

It must be remembered that calculations of Brandt \cite{brandt03}, which we have discussed above, are valid for conventional superconductors only. There are no reasons to believe that the vortex core structure should be the same in two-gap superconductors or in superconductors with nodes in the order parameter. Furthermore, one may assume that the influence of the vortex core region on the distribution of the magnetic induction should be even stronger than in the case of conventional superconductors. This is why, if calculations of \cite{brandt03} or any other calculations based on the conventional GL theory are used for the analysis of $\mu$SR data collected in different magnetic fields, it would produce an unphysical $\lambda(H)$ dependence. Although this result does not mean any special behavior of the magnetic field penetration depth, it should be considered as interesting. Indeed, if the conventional GL theory cannot describe the results of $\mu$SR experiments and all other possibilities for this disagreement are excluded,\footnote{For instance, the traditional analysis cannot be used in the case of polycrystalline samples of anisotropic superconductors.} one may conclude that this superconductor is unconventional. 

Superconductors with $d$-pairing as well as two-gap superconductors are more complex than conventional ones. For instance, two lengths $\xi_0$ and $\lambda_0$ are insufficient for their characterization and some additional information is needed. At present, there is no experimentally proven theory of the mixed state in unconventional superconductors. In this sense, magnetic field dependences of muon relaxation rates cannot be interpreted quantitatively without some additional assumptions. At the same time, one can try to obtain $F(B/B_{c2})$ experimentally in order to compare results for different superconducting materials. Unfortunately, concerning high-$T_c$ superconductors, the $H_{c2}(T)$ curves are not yet reliably established.

Interesting theoretical approaches for interpretation of the $\mu$SR experiments in the case of $d$-pairing was developed in \cite{ishi1,ishi2,ishi3,ishi4,affl97,affl97-2,affl98,affl00}. In  \cite{affl97,affl97-2,affl98,affl00} was convincingly argued that because of the nodes of the order parameter, the electrodynamics of the mixed state becomes nonlocal. This nonlocality effectively increases the vortex core radius and changes the distribution of the magnetic induction around vortices (see Fig. 6 of Ref. \cite{affl98}). If this effect is not taken into account, the magnetic field penetration depth evaluated from $\mu$SR experiments will be overestimated and magnetic field dependent.  The distortion of the results is very clearly demonstrated in Fig. 4 of Ref. \cite{affl98}. In order to correct the results, the function $\lambda_{eff}(B)/\lambda_0$ was introduced \cite{affl98}. Using this function, which is an analog of $F(B/B_{c2})$, one can evaluate the magnetic field penetration depth $\lambda_0$.

At the same time, in high $\kappa$ superconductors and at low magnetic inductions, the total volume of vortex cores is small and contribution of vortex cores cannot considerably change the muon signal. In this case, one may use $F = 0.061\Phi_o$ for evaluation of $\lambda_0$ also in unconventional superconductors. Because $\lambda \sim 1/\sqrt{\sigma}$, the resulting error is not expected to be big. This means that, if an experimental $\sigma(H)$ dependence is available, extrapolation of  $\sigma(H)$ (or $1/\sqrt{\sigma}$) to $H = 0$ gives more reliable values of $\lambda_0$. 

\section{Analysis of experimental results}

In this section, in order to simplify notation, we shall use $\lambda$ and $\xi$ without indexes, having in mind the magnetic field penetration depth and the superconducting coherence length as they are introduced in the GL theory.

\subsection{RbOs$_2$O$_6$, Cd$_2$Re$_2$O$_7$, PrOs$_4$Sb$_{12}$.}

\begin{figure}[!h]
\begin{center}
  \epsfxsize=0.8\columnwidth \epsfbox {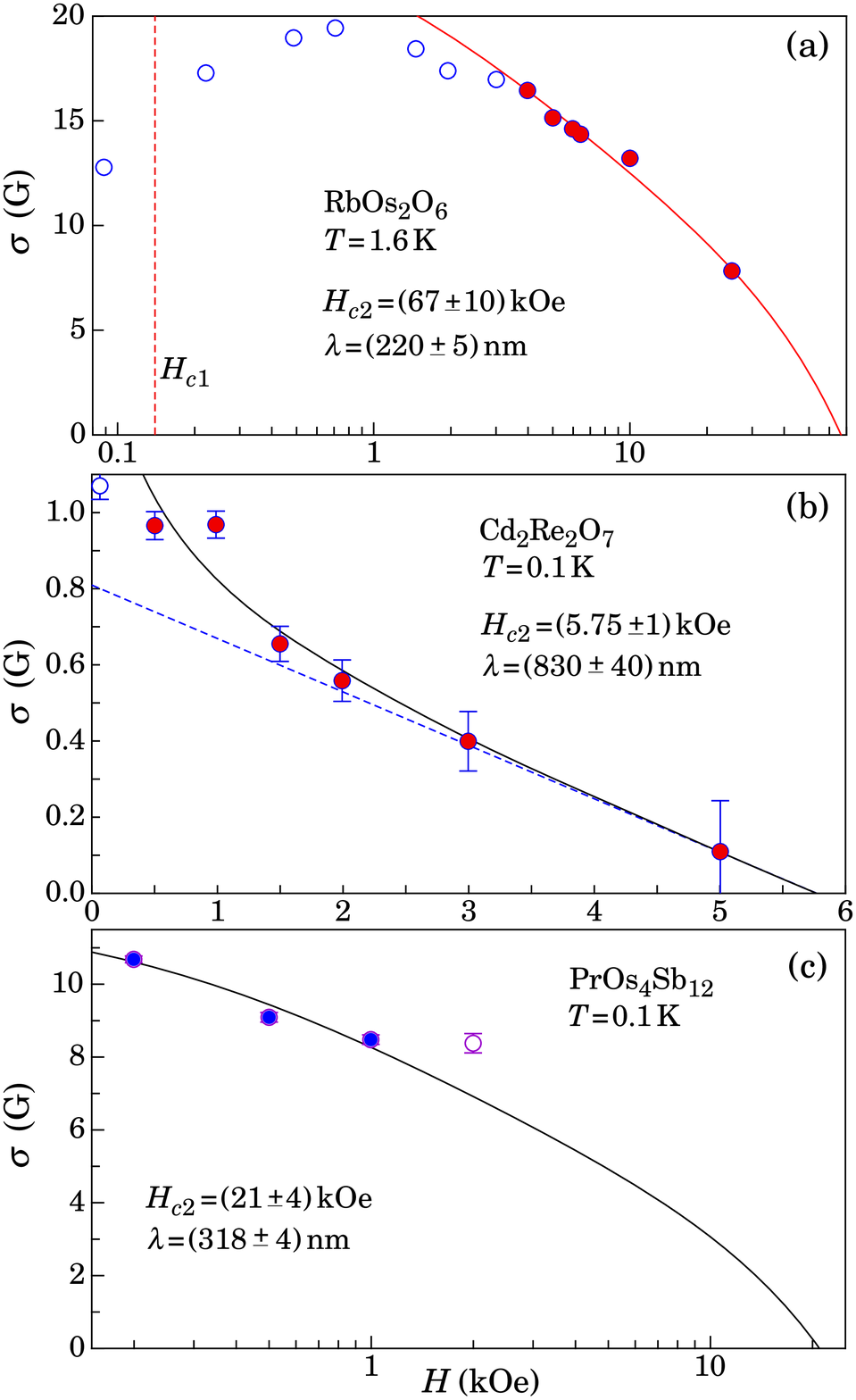}
  \caption{$\sigma(H)$ data for three different superconducting compounds. The solid lines represent the theoretical $\sigma(B/B_{c2})$ curve fitted to data-points. Only the data marked by closed symbols were used for fitting. The resulting values of $\lambda$ and $H_{c2}$ are indicated in the figure. (a) RbOs$_2$O$_6$ sample studied in \cite{khas05}. The vertical dashed line indicates the value of $H_{c1}$. (b) Cd$_2$Re$_2$O$_7$ sample studied in \cite{son02}. The dashed line represents a linear approximation to a high field part of the theoretical $\sigma(H)$ curve. (c) PrOs$_4$Sb$_{12}$ studied in \cite{pros1}.}
 \end{center}
\end{figure}
Experimental $\sigma(H)$ data for a polycrystalline sample of RbOs$_2$O$_6$ are shown in Fig. 2(a). Because RbOs$_2$O$_6$ is an isotropic superconductor, using of such samples is justified. This sample was investigated in \cite{khas05} and experimental data were analyzed by employing of an interpolation formula proposed in \cite{brandt88}. Because this formula deviates significantly from more accurate numerical calculations (see Fig. 1), we reanalyze these data using calculations of Ref. \cite{brandt03}. As may be seen in Fig. 2(a), experimental data-points for $H > 2$ kOe can very well be fitted by the theoretical $\sigma(B/B_{c2})$ curve. This fit results in $H_{c2} = (67 \pm 10)$ kOe and $\lambda = (220 \pm 5)$ nm (the value of $\lambda = 260$ nm was obtained in the original publiction). Because the value of $H_{c2}$ is obtained by the extrapolation of the $\sigma(H)$ curve to $\sigma = 0$, the corresponding error margins  are large. It is important to emphasize that $H_{c2}$, evaluated in such a way, is in agreement with the $H_{c2}(T)$ curve presented in \cite{khas05}. This agreement together with sufficiently high quality of fitting strongly supports our analysis.

We have chosen a high field part of the experimental $\sigma(H)$ curve for the analysis because in higher magnetic fields $F(B/B_{c2})$ is independent of $\kappa$ (see Fig. 1). In principle, analyzing the low field part of the curve, the value of $\kappa$ may straightforwardly be evaluated. This, however, is not always feasible. As was already mentioned, the correct parameter is not $H$ but the magnetic induction $B$. The value of $B$ determines intervortex distances and all other characteristics of the mixed state. In magnetic fields $H \gg H_{c1}$, the difference $(H - B) \ll H$ and one can use $H$ instead of $B$. In low fields, however, the equilibrium value of $B$ is considerably smaller than $H$ and the actual difference $(H - B)$ depends on pinning and on the demagnetizing factor of the sample. Furthermore, in low fields, the magnetic induction is nonuniform throughout the sample if its shape is not ellipsoidal. In polycrystalline samples, the situation complicates even further. Indeed, in such samples, some vortices may go along intergrain boundaries, which can significantly influence the magnetic induction distribution. This is the reason that we do not speculate on the low-field behavior of the $\sigma(H)$ curve.

Similar results for a Cd$_2$Re$_2$O$_7$ sample studied in \cite{son02} are shown in Fig. 2(b). For the reasons explained above, we disregard the lowest field data-point. Again in this case, data can be very well fitted with the GL theory, providing $H_{c2} = (5.75 \pm 1)$ kOe in agreement with the original data (see \cite{son02}). The value of $\lambda = (830 \pm 40)$ nm is also close to the result $\lambda = 750$ nm of Ref. \cite{son02}. We also note that approximation of experimental $\sigma(H)$ data-ponts by a linear dependence, as it was done in \cite{son02} and some other publications, is unjustified. As may be seen in Figs. 1 and 2(b), the theoretical $\sigma(B)$ curves are not at all linear.

Fig. 2(c) presents $\sigma(H)$ data for a heavy-fermion superconductor PrOs$_4$Sb$_{12}$ \cite{pros1}. We do not discuss here different features of this rather unusual superconductor but limit ourselves to one simple question whether the $\sigma(H)$ dependence for this compound can be described by the conventional GL theory. 

As may be seen in Fig. 2(c) (see also Fig. 4 of Ref. \cite{pros1}), the values of $\sigma(1$kOe) and $\sigma(2$kOe) practically coincide. It was assumed in Ref. \cite{pros1} that a change of vortex lattice symmetry or some other important changes of the vortex structure, which occur in magnetic fields above $H = 1$ kOe, may be responsible for such a behavior. This explanation seems to be plausible and we, as a precaution, do not use the highest field data-point in the analysis. 

The solid line in Fig. 2(c) represents the best fit of the theoretical $\sigma(H)$ curve to the data collected in magnetic fields $0.2$ kOe $\le H \le 1$ kOe. Quite amazingly, the resulting value of $H_{c2} = 21$ kOe practically coincides with $H_{c2} = 22.2$ kOe obtained in Ref. \cite{pros3} from resistivity measurements. The value of $\lambda = (318 \pm 4)$ nm is also close to the result $\lambda = 290$ nm of the original publication.

\begin{table}[h]
\caption{ }
\begin{tabular}{lcccccc}
\multicolumn{1}{c}{  } &
\multicolumn{1}{c}{RbOs$_2$O$_6$} (1.6K) &
\multicolumn{1}{c}{Cd$_2$Re$_2$O$_7$} (0.1K) &
\multicolumn{1}{c}{PrOs$_4$Sb$_{12}$} (0.1K) &   \\

$H_{c2}$ (kOe) & $67 \pm 10$ & $5.75 \pm 1$ & $21.4 \pm 4$   \\
$\kappa$ & $34 \pm 1$ & $35 \pm 3$  & $25.5 \pm 2.5$  \\ 
$\lambda$ (nm) & $220 \pm 5$ & $830 \pm 40$ & $318 \pm 4$   \\
$\xi$ (nm) & $7 \pm 0.4$ & $24 \pm 3.5$ & $12.5 \pm 1.5$ &  \\

\end{tabular}
\end{table}
The main characteristics of the superconducting compounds, resulting from our analysis of the $\mu$SR data published in Refs. \cite{khas05,son02,pros1}, are listed in Table 1. We emphasize that all parameters were evaluated by fitting of $\sigma(H)$ data-points with the theoretical $\sigma(B/B_{c2})$ dependence calculated in \cite{brandt03}. In all cases, the values of $H_{c2}$ practically coincide with results of independent measurements. 

\subsection{CeRu$_2$\cite{kado}.}

Calculation of $\sigma(T,H)$ considered above is not the only way of analysis of $\mu$SR experiments in the mixed state of type-II superconductors. A different method was employed in Refs. \cite{son97,son97-3,son99,ohi,son03,son97-2,son04,son05,son06}. In this approach, the distribution of local fields (the Fourier transform of the muon precession signal) $P(B)$ was directly analyzed by comparing with corresponding theoretical calculations. In real experiments, however, the $P(B)$ line is usually different from theoretical predictions. This difference is expected. Indeed, the calculations are made for a perfect sample and for a perfect vortex lattice. All imperfections, which cannot be avoided in experiments, influence the $P(B)$ curves. This is why, in order to approximate experimental data with theoretical calculations, some gaussian smearing factor is introduced. In such a way, satisfactory agreement between the theory and experiments can be achieved. This is justified if it is {\it a priori} known that the spatial distribution of the magnetic induction around vortex lines is in agreement with the theory, which is used for the calculations. If it is not the case, introducing of additional Gaussian relaxation may mask the disagreement and provide misleading results. 

For some type-II superconductors, the $P(B)$ curves for different values of the applied magnetic field are available in the literature. This allows to calculate $\sigma(H)$ and to employ the same kind of the analysis as was used above. Below we present the results of such analysis for single crystals of CeRu$_2$ and vanadium. 

\begin{figure}[h]
\begin{center}
  \epsfxsize=0.8\columnwidth \epsfbox {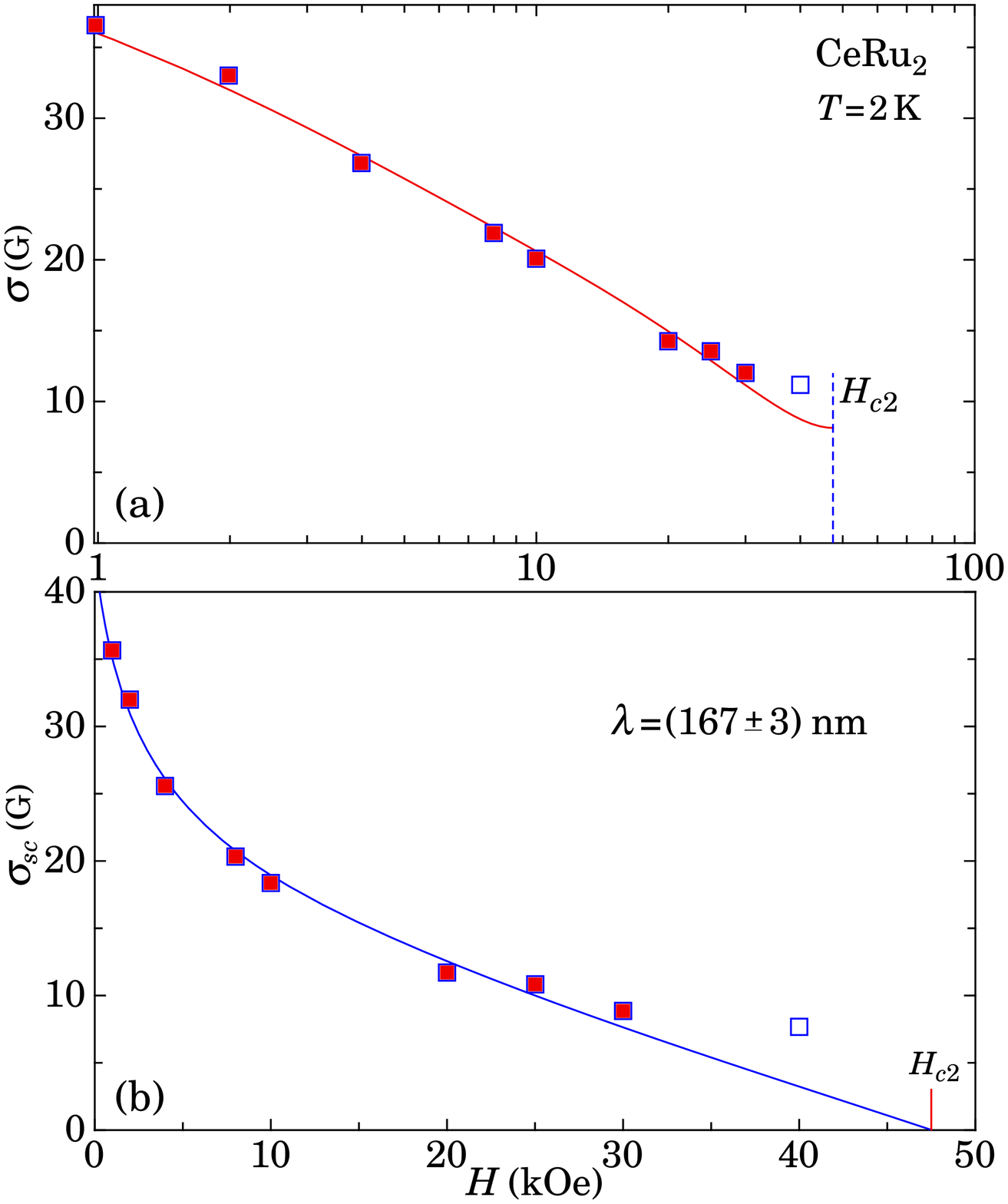}
  \caption{$\sigma$ (upper panel) and $\sigma_{sc} = \sqrt{\sigma^2 - \sigma^2_{bg}}$ (lower panel) versus $H$ for a CeRu$_2$ sample studied in \cite{kado}. The solid lines represent the theoretical curves calculated as explained in the text. The chosen value of $H_{c2} = 47.5$ kOe is indicated in the figures by vertical lines. Only data-points shown by closed symbols were used for evaluation of $\sigma_{bg}$ and $\lambda$.}
 \end{center}
\end{figure}
Fig. 3(a) shows $\sigma$ as a function of $H$ for a CeRu$_2$ sample experimentally investigated in Ref. \cite{kado}. The difference to the results displayed in Fig. 2 is that broadening of the $P(B)$ line resulting from other sources of field inhomogeneity was not accounted for. In this case $\sigma$, evaluated from $\mu$SR experiments may be written as 
\begin{equation}
\sigma = \sqrt{\sigma^2_{sc} + \sigma^2_{bg}},
\end{equation}
where $\sigma_{sc}$ and $\sigma_{bg}$ are the mixed state and background contributions, respectively. As may be seen in \cite{kado}, $\sigma_{bg}$ is not small and cannot be evaluated from the data presented in the publication with sufficient accuracy. This is the reason that we introduce $\sigma_{bg}$ as an additional adjustable parameter. Because experimental data are insufficient for evaluation of $\lambda$, $H_{c2}$ and $\sigma_{bg}$ together, we take the value of $H_{c2}$ from the original publication.\footnote{There is some confusion in Ref. \cite{kado}. While Fig. 2 provides $H_{c2} = 50$ kOe, the value of $H_{c2}$ evaluated from Fig. 1 is closer to 45 kOe. Taking into account that the resulting $\lambda$ is not very sensitive to some variation of the assumed $H_{c2}$ value, we have chosen $H_{c2} = 47.5$ kOe for the analysis.} We also note that $H = 40$ kOe is the only data point corresponding to the peak-effect region (see Fig. 1 in \cite{kado}). Because the origin of this effect is not yet established, we exclude the corresponding data point from the analysis. 

As may be seen in Fig. 3(a), all data-points for $H \le 30$ kOe can be fairly well fitted by the theory, providing $\lambda = (167 \pm 3)$ nm and $\sigma_{bg} = (8.4 \pm 1)$ G. The magnetic field dependence of $\sigma_{sc} = \sqrt{\sigma^2 - \sigma^2_{bg}}$ is shown in Fig. 3(b). 

\begin{figure}[!h]
\begin{center}
  \epsfxsize=0.8\columnwidth \epsfbox {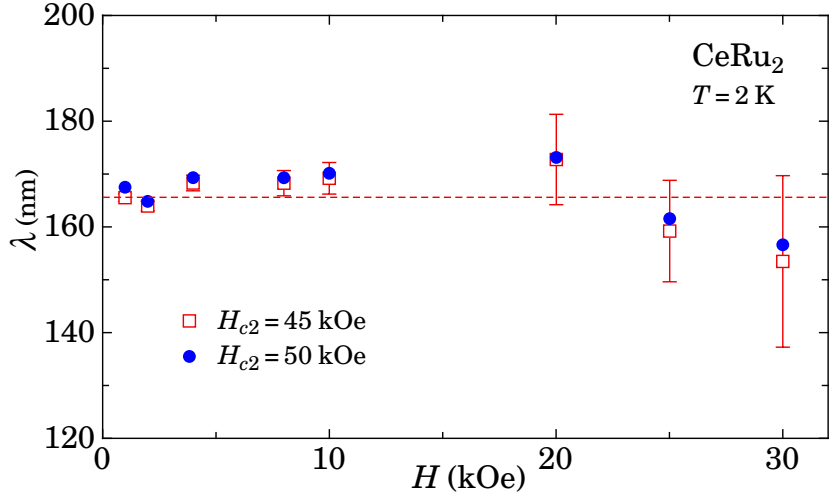}
  \caption{$\lambda$ as a function of $H$ calculated assuming $H_{c2} = 45$ kOe and $H_{c2} = 50$ kOe.}
 \end{center}
\end{figure}
Using the $\sigma_{sc}(H)$ plot presented in Fig. 3(b), we can calculate $\lambda$ for each of the data-points. Such calculations were made for two different values of $H_{c2}$ and they are presented in Fig. 4. As was expected, the absolute value of $\lambda$ is practically independent of the chosen value of $H_{c2}$. One can also see that, contrary to claims of Ref. \cite{kado}, there is no any noticeable dependence of $\lambda$ on $H$. 

At the same time, the value of $\sigma(40$kOe$)$ deviates quite significantly from the theoretical curve (see Fig. 3).\footnote{$\sigma(40$kOe$)$ is larger than the corresponding theoretical value. The higher $\sigma$ means smaller $\lambda$. This conclusion is just opposite to that made in the original publication.} If this deviation is not an experimental error, it means that the distribution of the magnetic induction in the case of the peak-effect is rather different in comparison with the conventional mixed state. However, one should be extremely careful with such conclusions. In the case of the peak-effect, the value of $\sigma$ is rather sensitive even to insignificant variations of $H$ (see Fig. 13(d) in \cite{kado}). In this situation, $\sim 10^{-5}H$ change of the external field may explain the difference between $\sigma(40$kOe$)$ and the theoretical curve.

\subsection{Vanadium single crystal \cite{son06}}

We discuss experimental data of Ref. \cite{son06} in some detail in order to demonstrate general problems of interpretation of $\mu$SR experiments in the case of low-$\kappa$ superconductors. We also discuss some typical errors that can be found in the literature.

\begin{figure}[h]
\begin{center}
  \epsfxsize=0.8\columnwidth \epsfbox {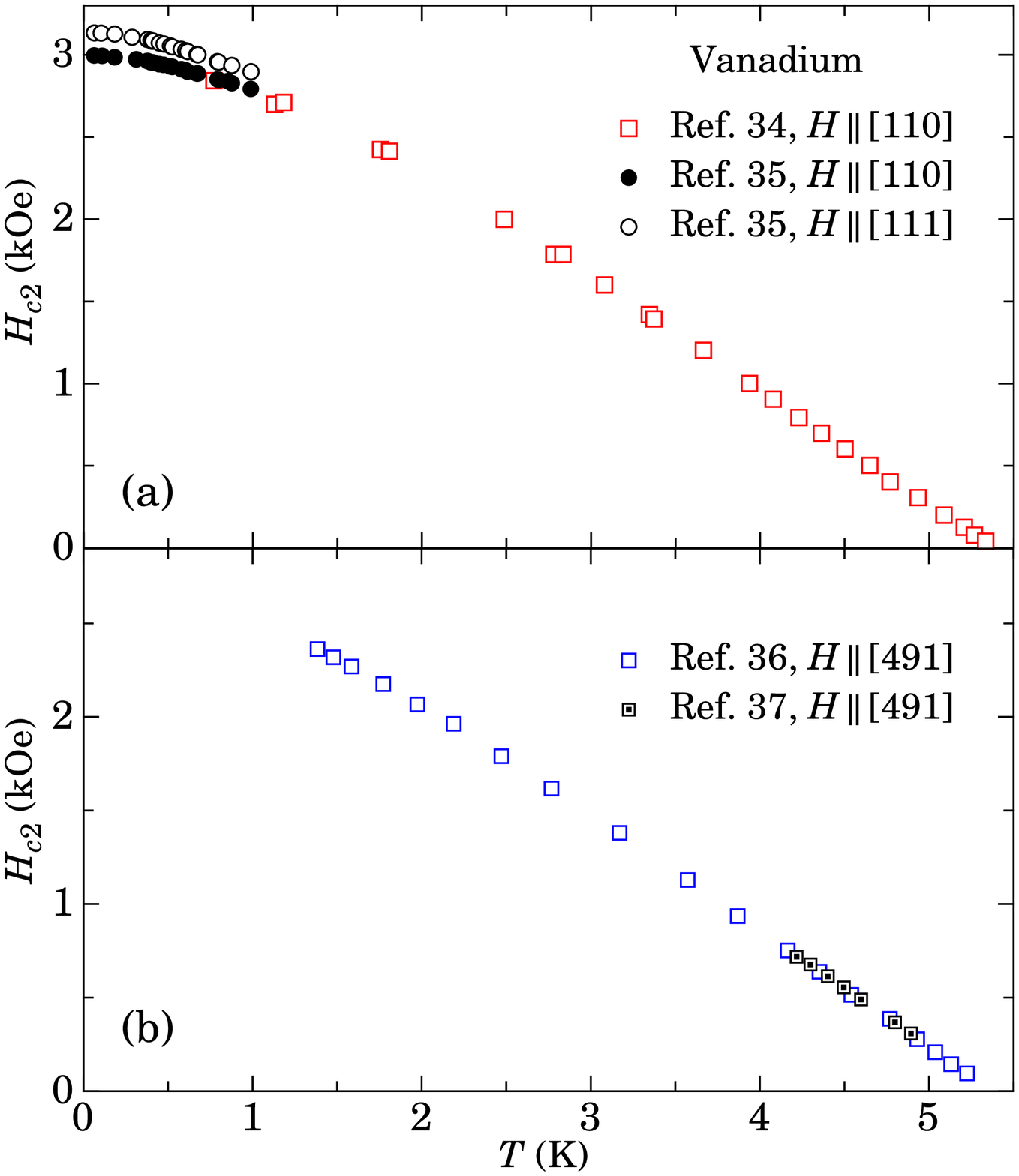}
  \caption{$H_{c2}(T)$ for three different orientations of vanadium
   single crystals}
 \end{center}
\end{figure}
Vanadium is one of the very few pure metals, which displays type-II superconductivity at all temperatures. Superconducting characteristics of vanadium have been rather well investigated (see, for instance, \cite{keesom,william,sekula,boykin}). Although vanadium has a cubic (bcc) structure, $H_{c2}$ depends on the orientation of the applied magnetic field \cite{william}. According to \cite{william}, the value of $H_{c2}$ along [111] direction is approximately 10\% higher than that for [001]. 

\begin{figure}[h]
\begin{center}
  \epsfxsize=0.8\columnwidth \epsfbox {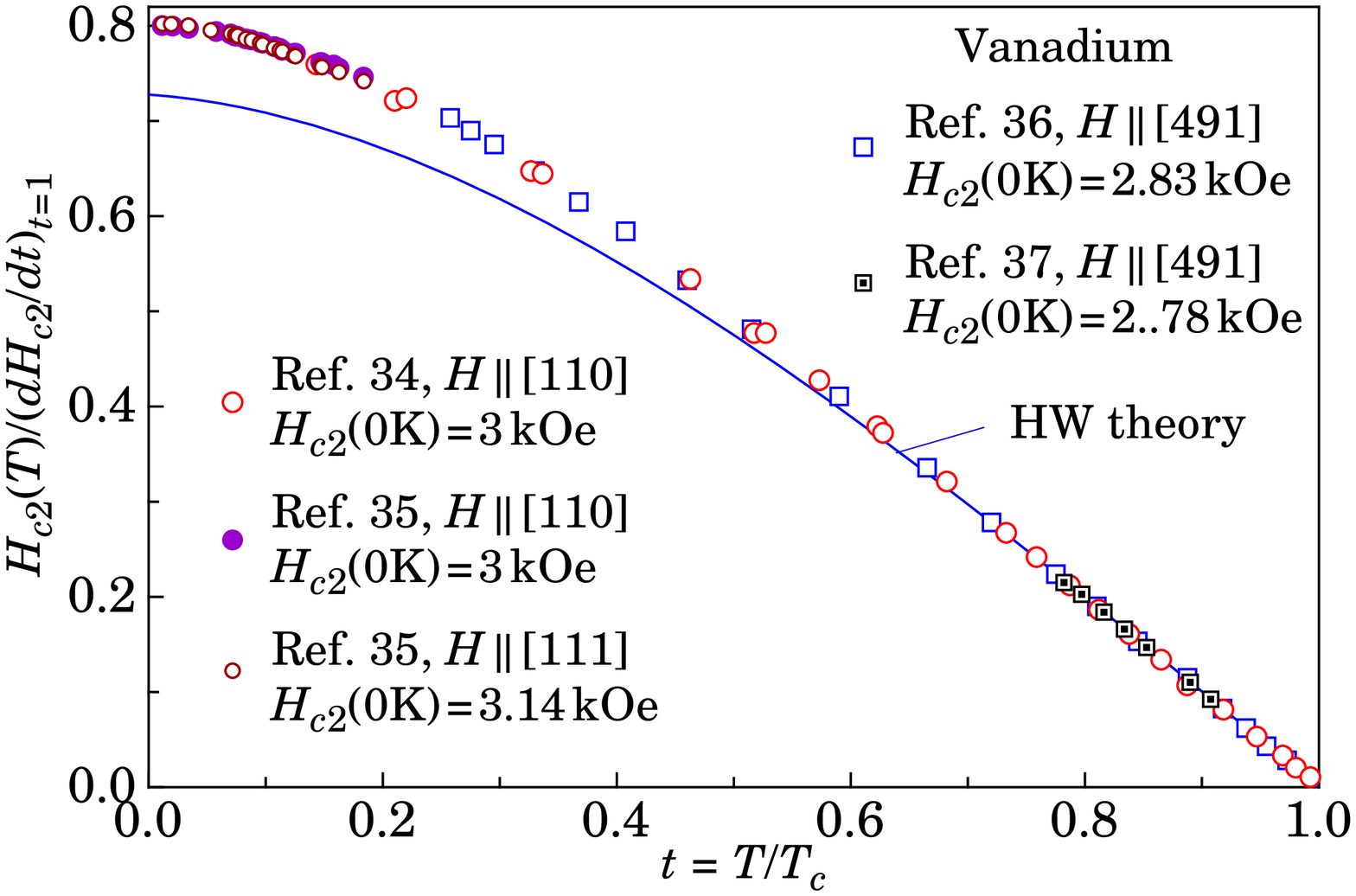}
  \caption{The normalized upper critical field of vanadium single crystals as a function of $t = T/T_c$.  The solid line represents the HW theory \cite{hw}.}
 \end{center}
\end{figure}
The dependences $H_{c2}(T)$ for three different orientations are shown in Fig. 5. As may be seen, results of different studies are in excellent agreement. The value of $T_c$ may be evaluated as $T_c = (5.40 \pm 0.05)$ K \cite{keesom,william,sekula,boykin}. While $H_{c2}$ is orientation dependent, its normalized temperature dependence is practically universal \cite{william}. This is illustrated in Fig. 6 where $H_{c2}(T)/(T_cdH_{c2}/dT)_{T=T_c}$ is plotted versus $T/T_c$. The results of different works, presented  in such a way, nicely collapse onto a single curve. We note that the temperature dependence of $H_{c2}$ is somewhat different from predictions of Helfand-Werthamer (HW) theory \cite{hw}.

Vanadium is a low-$\kappa$ material with $\kappa(0$K$) =1.5$ for a pure sample investigated in \cite{keesom}. This circumstance adds some peculiarities to the mixed state and its description. First, the condition $H \ll H_{c2}$ is not satisfied even in magnetic fields down to $H_{c1}$, i.e., the London approach, in which vortices are considered as independent, is inapplicable in the entire range of magnetic fields. In this situation, the actual magnetic induction distribution in the sample strongly depends on spatial variations of the order parameter, and the accuracy of the corresponding calculations plays a crucial role.  Second, if $\kappa \sim 1$, the condition $\lambda \gg \xi_P$ ($\xi_P = 0.74\xi(0$K) is the Pippard coherence length) is not satisfied at low temperatures and electrodynamics become nonlocal, i.e., quantitative applicability of the GL theory at $T \ll T_c$ is questionable. Furthermore, the results of Refs. \cite{keesom,william,sekula} clearly demonstrate that superconducting properties of vanadium at $T \ll T_c$ cannot be described by the GL theory  and a more complex approach is necessary. At the same time, as we argue below, experimental $\sigma(H)$ curves are close to theoretical predictions of Brandt \cite{brandt03} and can be used for evaluation of the magnetic field penetration depth. 

\begin{figure}[h]
\begin{center}
  \epsfxsize=0.8\columnwidth \epsfbox {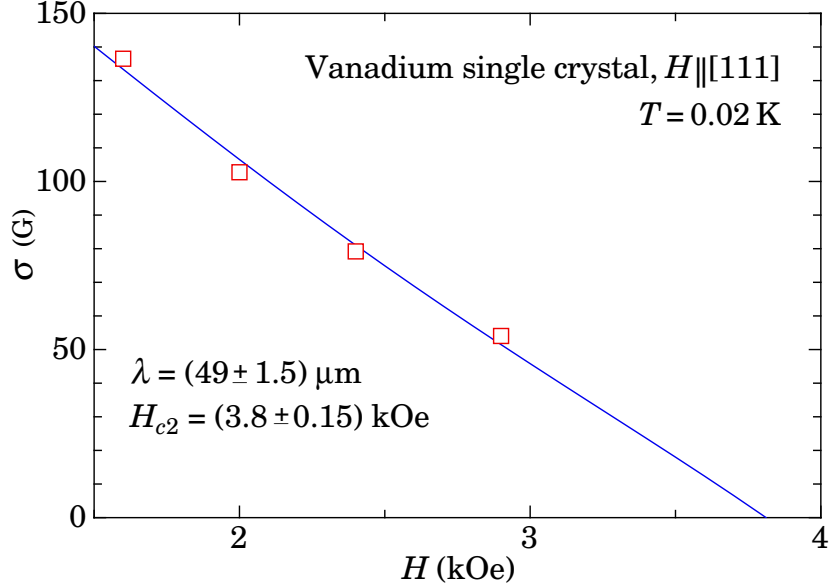}
  \caption{(a) $\sigma$ versus $H$ for a vanadium single crystal studied in \cite{son06}. The solid line represent a fit with the Brandt theory \cite{brandt03}. The resulting values of $\lambda$ and $H_{c2}$ are indicated in the figure. }
 \end{center}
\end{figure}
The values of $\sigma$ were calculated using the $P(B)$ curves presented in Ref. \cite{son06}. Clearly visible peaks arising from muons stopped outside the sample were approximated by Gaussians and subtracted from the data. The resulting values of $\sigma$ are plotted in Fig. 7(a).\footnote{Because the demagnetizing factor of the sample is close to 1, one can safely assume $H = B$ for all considered magnetic fields.} Our analysis gives $\lambda = (49 \pm 1.5)$ nm and $H_{c2} = (3.8 \pm 0.15)$ kOe. The estimation of $\lambda$ is in very good agreement, with $\lambda = 50$ nm, which may be calculated from $H_{c2}(T)$ and $\kappa(T)$ curves experimentally measured in \cite{keesom}. 

The value of $H_{c2}$ evaluated above is just 10\% below of $H_{c2} = 4.2$ kOe provided in the original publication \cite{son06}. We note that there are no estimations of experimental uncertainty for $H_{c2}$ in \cite{son06}. One can assume that $H_{c2}(0.02$K) was obtained by extrapolation of higher temperature data and the corresponding error margins are considerable. We also note that $H_{c2}(0.02$K$) = 4.2$ kOe is well above earlier results (see Fig. 5). Partly this difference may be explained by the fact that the sample that we are discussing here was substantially less pure than those of Refs. \cite{keesom,william,sekula,boykin}. However, such a significant increase of $H_{c2}$ seems to be unlikely. Furthermore, as we show below, the value of  $H_{c2}$, evaluated by the analysis of the temperature dependence of muon relaxation rates, agrees better with the estimate of $H_{c2}$ presented in Fig. 7 than with the value given in \cite{son06}.

\begin{figure}[h]
\begin{center}
  \epsfxsize=0.8\columnwidth \epsfbox {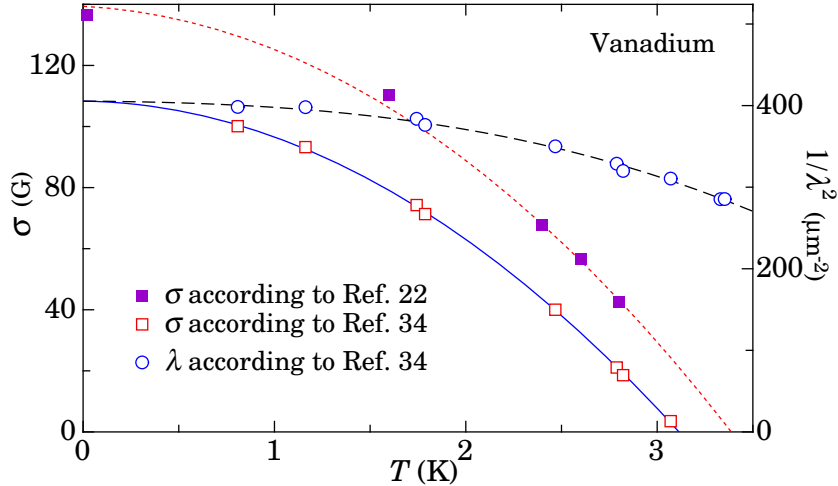}
  \caption{Temperature dependence of $\sigma(1.6$kOe) for a vanadium single crystal studied in  \cite{son06}. The temperature dependences of $\sigma$ and $\lambda^{-2}$ (right $y$-axis), evaluated from experimental data of \cite{keesom}, are shown for comparison. The  solid and the dashed lines are the guides to the eye. The dotted line ($f'(T)$) is obtained by scaling of the solid line ($f(T)$), i.e., $f'(T) = 1.29f(1.09T)$.}
 \end{center}
\end{figure}
As was already mentioned, experimental results presented in \cite{keesom} allow for evaluation both $H_{c2}(T)$ and $\lambda(T)$ dependences. Using these data, we can also obtain the expected value of $\sigma$ for any magnetic field and temperature. Such results for $H = 1.6$ kOe are plotted in Fig. 8 for comparison with the $\mu$SR data of Ref. \cite{son06}. As may be seen in Fig. 8, the two $\sigma(T)$ curves are similar. In order to emphasize this similarity, we approximate both data sets by the same functional dependence (see Fig. 8). Considering the results presented in Fig. 8, one can conclude that the sample investigated in \cite{son06} has indeed a somewhat higher $H_{c2}$. The value of $T_c(H)$ may straightforwardly be evaluated from $\sigma(T)$ data as the value of $T$, at which $\sigma$ vanishes. Such estimate gives $H_{c2} = 1.6$ kOe at $T = 3.38$ K.\footnote{This is well below the value of $T_c(1.6$kOe$) = 3.65$ K provided in \cite{son06}.} Using this value of $H_{c2}(3.38$K) and the normalized $H_{c2}(T)$ curve presented in Fig. 6, we can evaluate $H_{c2}(0) = (3.4 \pm 0.25)$ kOe, which is in reasonable agreement with the estimate made from the analysis of $\sigma(H)$ data (see Fig. 7).\footnote{We use $T_c = 5.4$ K, as it follows from earlier measurements (the same value is provided in reference data of Goodfellow Ltd.), assuming that $T_c = 5.2$ K given in \cite{son06}, is a misprint. If, however, we except $T_c = 5.2$ K, $H_{c2}(0) = (3.65 \pm 0.2)$ kOe, the value practically coinciding with the result of Fig. 7.}

Our results presented in this section are rather different from the conclusions of Ref. \cite{son06}. First and foremost, as it is clearly demonstrated in Fig. 7, the magnetic field dependence of $\sigma$ is very close to the result of the GL theory. Our value of $\lambda$ for $T = 0$, which is in excellent agreement with measurements of Ref. \cite{keesom}, is about 1.5 times smaller than the result of \cite{son06} for $H = 1.6$ kOe. The temperature dependence of $1/\lambda^2$ calculated according to Ref. \cite{keesom} is also plotted in Fig. 8. As may be seen, while $\sigma$ vanishes at $T = 3.12$ K, the value of $\lambda(3.12$K) remains almost the same as at $T = 0$. In other words, $\sigma$ vanishes at $T_c(H)$ not because of the divergence of $\lambda$ but because the coefficient $F$ in Eq. (7) vanishes at this temperature.  

It seems important to emphasize that Fig. 10 of Ref. \cite{son06} is based on an obvious misunderstanding.\footnote{A similar plot one can also find in \cite{khas05}} There exists no theory that predicts divergence of $\lambda$ at $T_c(H)$. The reference on theoretical calculations of M\"uhlschlegel \cite{muhl}, given in \cite{son06}, is misleading. Indeed, the thermodynamical consideration of Ref. \cite{muhl} is based on the fact that the difference between free energies of the normal and the superconducting states per unit of volume can be written as $H_c^2/8\pi$. The same difference can also be written as $n_{cp} \Delta$ where $n_{cp}$ is be the density of Cooper pairs and $\Delta$ is the equilibrium (zero-field) superconducting energy gap. Using this, one obtains $n_{cp}(T)$ and $\lambda(T)$. Nothing in this consideration can be used to justify Fig. 10 of Ref. \cite{son06}. We also note that the temperature variation of $H_{c2}$ should be taken into account if the temperature dependence of $\lambda$ is evaluated from measurements in fixed magnetic fields. It can be neglected only if the condition $H \ll H_{c2}$ is satisfied at all temperatures. 

In fact, good agreement with the theory, demonstrated in Figs. 7 and 8, is rather surprising. As was already mentioned, it is well established that vanadium does not obey the GL theory at $T \ll T_c$ \cite{keesom,william,sekula}.The most probable is that the distribution of the magnetic induction in the sample ($P(B)$) is different from theoretical predictions, while $\sigma$, as a more integral characteristic of this distribution, remains practically the same. This assumption can also explain the difference between our results and those of Ref. \cite{son06}. Indeed, the analysis of $P(B)$ functions, carried out in \cite{son06}, resulted in an unphysical magnetic field dependence of $\lambda$, which clearly demonstrates the inapplicability of the GL theory to this analysis. 

It is important to underline that, although the distribution of the magnetic induction in the sample cannot be described by the GL theory in low-$\kappa$ type-II superconductors at  $T \ll T_c$, the $\sigma(H)$ curves can still be used for evaluation of the magnetic field penetration depth, as it is proven by a very close agreement between our value of $\lambda(0$K$) = 49 \pm 1.5$ nm and 50 nm calculated from results of Ref. \cite{keesom}. We also note that at temperatures closer to $T_c$ the GL theory should be applicable and both analyses should result in the same $\lambda$ values.

\subsection{YNi$_2$B$_2$C \cite{ohi}}

In order to demonstrate that in some cases the GL theory cannot describe $\mu$SR data, we consider a study of a borocarbide superconductor YNi$_2$B$_2$C \cite{ohi}. Rare-earth nickel borocarbide superconductors attracted a lot of attention during the past decade. Already the very first  studies of YNi$_2$B$_2$C demonstrated a pronounced positive curvature of the $H_{c2}(T)$ curve, indicating unconventional superconductivity \cite{YNiHc2,YNiHc2-2}. Similar conclusions were made from specific heat data \cite{YNiHc2,YNiHc2-2}. Although YNi$_2$B$_2$C has been extensively studied, the nature of this unconventionality is still under discussion. While Refs. \cite{YNi-n1,YNi-n2,YNi-n3,YNi-n4,YNi-n5,YNi-n6} provide evidences of point nodes in the superconducting gap function, other works point out on multiband superconductivity \cite{YNi-2g1,YNi-2g2,YNi-2g3,YNi-2g4}. The distinction between these two possibilities is sometimes difficult to make. For instance, as was recently demonstrated, specific heat data may be fitted equally well by nodal and two-gap models \cite{huang}.
\begin{figure}[h]
\begin{center}
  \epsfxsize=0.8\columnwidth \epsfbox {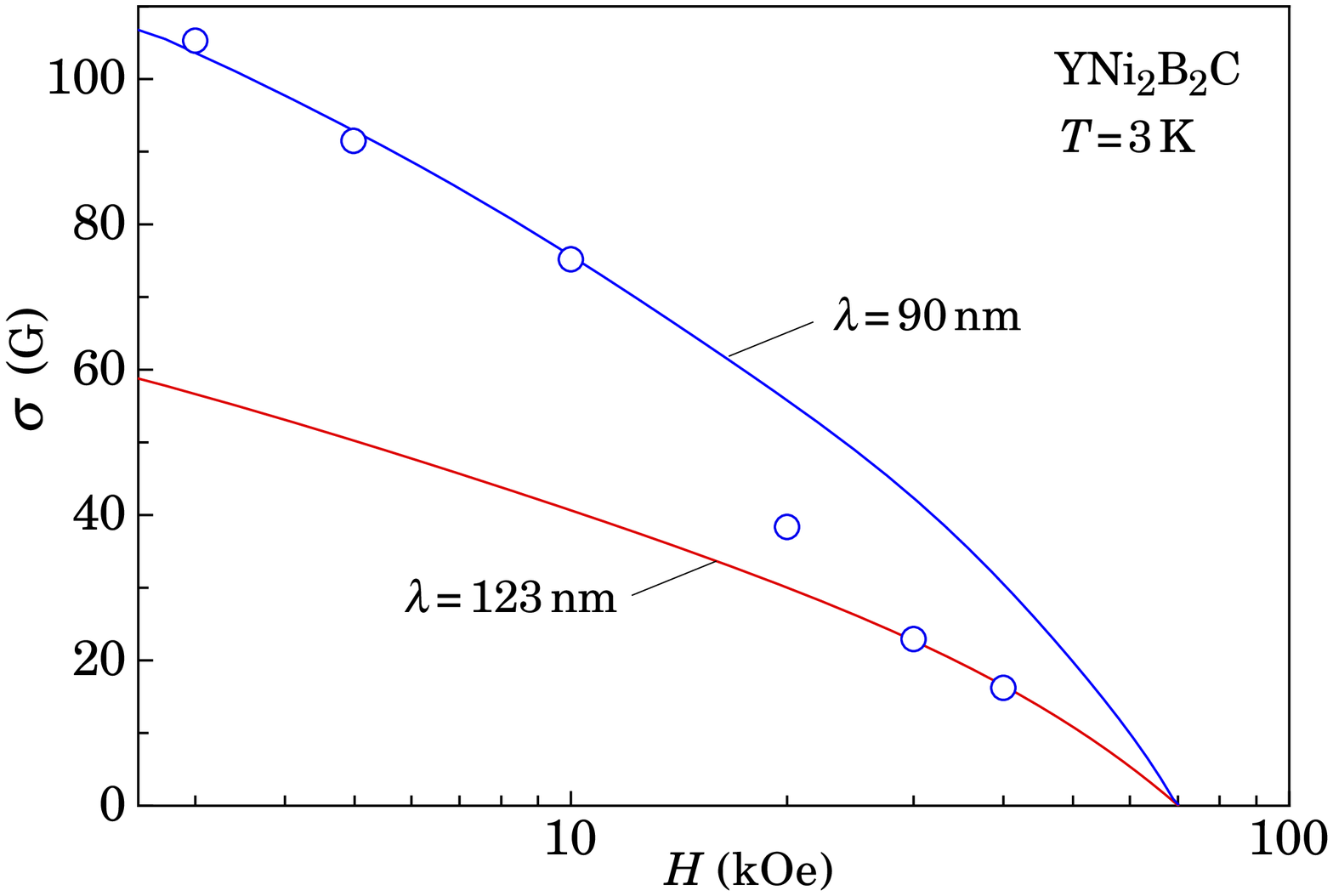}
  \caption{$\sigma$ as a function of $H$ for a YNi$_2$B$_2$C sample studied in Ref. \cite{ohi}. The solid lines are the theoretical $\sigma(H)$ dependencies calculated for $H_{c2} = 70$ kOe and for two different ¤values of $\lambda$. }
 \end{center}
\end{figure}

Experimental results of Ref. \cite{ohi} are plotted in Fig. 9 as $\sigma$ versus $H$. The value of $H_{c2}(3$K$) = 70$ kOe for this particular crystal is given in \cite{ohi}.  As may clearly be seen in Fig. 9, $\sigma(H)$ data-points cannot be fitted with the theory if the entire range of magnetic fields is considered. Because there are sufficient experimental evidences that YNi$_2$B$_2$C is an unconventional superconductor \cite{YNi-n1,YNi-n2,YNi-n3,YNi-n4,YNi-n5,YNi-n6,YNi-2g1,YNi-2g2,YNi-2g3,YNi-2g4,huang}, disagreement between the GL theory and experimental data is expected. We also note that in YNi$_2$B$_2$C a transition from a triangular to a square vortex lattice was observed \cite{esk,lev}. However, because this transition occurs in lower magnetic fields, it cannot have any influence on $\mu$SR data presented in Fig. 9.\footnote{As was demonstrated in \cite{ishi4}, although the magnetic induction distributions for square and triangular lattices are quite different, $\sigma(H)$ remains practically the same in both cases.}

While the totality of data cannot be fitted with the theory, both high-field and low-fild results may amazingly well be approximated with two different theoretical curves, corresponding to  two different $\lambda$ values (Fig. 9). Unfortunately, insufficient number of data-points does not allow to make unambiguous conclusions on this matter, however, if this behavior will be confirmed by a more detailed study, it may be considered as a rather interesting result, indicating two gap superconductivity.

In low magnetic fields $H \ll H_{c2}$, most of muons stopped outside vortex cores, i.e., the magnetic induction distribution in vortex core regions is not very important for $\mu$SR data. In this case, the difference between conventional and two gap superconductors should not be significant and the resulting $\sigma(H)$ curves can be close in these two cases. 

In higher magnetic fields, as it was established in studies of MgB$_2$,  superconductivity in one of two bands is completely suppressed and the superconductor behaves itself as a one gape superconductor but with a smaller number of Cooper pairs \cite{zhito,mgb1,mgb2}. This can explain the fact that the two data points for $H \ge 30$ kOe follow a standard theoretical curve with a higher value of $\lambda$ (see Fig. 9).

The quantity $1/\lambda^2$ is proportional to the density of Cooper pairs. If two gap superconductivity is assumed, the values of the magnetic field penetration lengths, evaluated from low- and high-field data, allows evaluation of relative weights of two superconducting bands. Such estimate gives 54\% and 46\% for stronger and weaker gaps, respectively. These values are noticeably different from the result 71\% and 29\% obtained in Ref. \cite{huang}. At present, however, it is too early to discuss such differences. Two data points in the high-field range part of the curve (see Fig. 9) are clearly insufficient in order to make any definite conclusion about superconductivity in YNi$_2$B$_2$C.

\section{Conclusion.}

In this work, we applied numerical calculations of Brandt \cite{brandt03} for the analysis of $\mu$SR experiments carried out in the mixed state of several superconducting compounds. It turned out that this approach may serve as a very powerful tool for the interpretation of $\mu$SR experiments. If the magnetic field dependences of muon depolarization rates are available, not only $\lambda$ but also $H_{c2}$ can  reliably be evaluated. We show that in the most of considered cases the magnetic field dependences of $\sigma$ may very well be described by a single and temperature independent $\lambda$.

In contrast to approximate analytical models, Ref. \cite{brandt03} provides precise  numerical solutions of 2-dimensional GL equations for different values of $\kappa$ ($0.85 \le \kappa \le 200$) and for magnetic fields ranging from $H_{c1}$ to $H_{c2}$. Using these solutions, different characteristics of the mixed state, including the $\sigma(B/B_{c2})$ dependences for various $\kappa$ values, were calculated. As well as we are aware, these calculations provide the best available description of the magnetic induction distribution in the mixed state of conventional type-II superconductors. We also note that numerical calculations of $\sigma(H)$ are available since 1997 (see Fig. 3 in \cite{brandt97}). In spite of this, for some mysterious reasons, these theoretical  calculations have practically never been used for the analysis of $\mu$SR data.

We also argued that the magnetic field dependence of $\lambda$ can never be obtained from analyses of experimental data collected in the mixed state. Indeed, because the local value of $\lambda$ is inversely proportional to the absolute value of the superconducting order parameter, one cannot introduce any single value of $\lambda$ in the mixed state.

Calculations of Brandt \cite{brandt03} represent the conventional GL theory and their validity for the description of unconventional superconductors is questionable.  In fact, there are no reasons to believe that the conventional GL theory can quantitatively describe either two-gap superconductors or superconductors with nodes of the order parameter and one should expect disagreement between Brandt's theory and experimental results in the case of unconventional superconductors, as it is demonstrated in Fig. 9.

We demonstrated that in conventional superconductors, the results of $\mu$SR experiments may be used for  the evaluation of both $\lambda$ and $H_{c2}$.  If applicability of the conventional GL theory is questionable, the knowledge of $H_{c2}$ is of primary importance. Disagreement between the values of $H_{c2}$ resulting from $\mu$SR data and independent measurements may be considered as convincing evidence that this particular superconductor is unconventional. 

We are grateful to R. Khasanov for numerous and fruitful discussions.


\begin{thebibliography}{80}

\bibitem{GL} V. L. Ginzburg and  L. D.  Landau, Zh. Exp. Teor. Fiz. {\bf 20}, 1064 (1950); 
Collected Papers of  L. D. Landau, D. Ter Haar (Ed.) Pergamon Press, London, 1965.

\bibitem{abr} A. A. Abrikosov, Zh. Eksp. Teor. Fiz. {\bf 37}, 833 (1959) [Soviet Phys. JETP {\bf 10}, 593 (1960)].

\bibitem{brandt88} E. H. Brandt, Phys. Rev. B {\bf 37}, 2349 (1988).

\bibitem{brandt97} A. Yaouanc, P. Dalmas de R\'eotier, and E. H. Brandt, Phys. Rev. B {\bf 55}, 11107 (1997).

\bibitem{brandt03} E. H. Brandt, Phys. Rev. Lett. {\bf 78}, 2208 (1997).  E. H. Brandt, Phys. Rev. B {\bf 68}, 054506 (2003).

\bibitem{khas-l} R. Khasanov, I.L. Landau, C. Baines, F. La Mattina, A. Maisuradze, K. Togano, and H. Keller, Phys. Rev. B {\bf 73}, 214528 (2006).

\bibitem{ishi1} N. Enomoto, M. Ichioka, and K. Machida, J. Phys. Soc. of Jpn. {\bf 66}, 204 (1997).

\bibitem{ishi2} M. Ichioka, N. Enomoto, and K. Machida, J. Phys. Soc. of Jpn. {\bf 66}, 3928 (1997).

\bibitem{ishi3} M. Ichioka, A. Hasegawa, and K. Machida, Phys. Rev. B {\bf 59}, 184 (1999).

\bibitem{ishi4} M. Ichioka, A. Hasegawa, and K. Machida, Phys. Rev. B {\bf 59}, 8902 (1999).

\bibitem{son97} J. E. Sonier, R. F. Kiefl, J. H. Brewer, D. A. Bonn, S. R. Dunsiger, W. N. Hardy, Ruixing Liang, W. A. MacFarlane, T. M. Riseman, Jr., D. R. Noakes, and C. E. Stronach, Phys. Rev. B  {\bf 55}, 11789 (1997).

\bibitem{son97-3} J. E. Sonier, J. H. Brewer, R. F. Kiefl, D. A. Bonn, S. R. Dunsiger, W. N. Hardy, Ruixing Liang, W. A. MacFarlane, R. I. Miller, T. M. Riseman, D. R. Noakes, C. E. Stronach, and M. F. White, Jr., Phys. Rev. Lett.  {\bf 79}, 2875 (1997).

\bibitem{son99} J. E. Sonier,  J. H. Brewer, R. F. Kiefl, G. D. Morris, R. I. Miller, D. A. Bonn, J. Chakhalian, R. H. Heffner, W. N. Hardy, and R. Liang, Phys. Rev. Lett.  {\bf 83}, 4156 (1999).

\bibitem{ohi} K. Ohishi, K. Kakuta, J. Akimitsu, W. Higemoto, R. Kadono, R. I. Miller, A. Price, R. F. Kiefl, J. E. Sonier, M. Nohara, H. Suzuki, and H. Takagi, Physica B  {\bf 289-290}, 377 (2000). K. Ohishi, K. Kakuta, J. Akimitsu, W. Higemoto, R. Kadono, J. E. Sonier, A. Price, R. I. Miller, R. F. Kiefl, M. Nohara, H. Suzuki, and H. Takagi, Phys. Rev. B  {\bf 65}, 140505 (2002).

\bibitem{son03} R. I. Miller, R. F. Kiefl, J. H. Brewer, J. C. Chakhalian, S. Dunsiger, A. N. Price, D. A. Bonn, W. H. Hardy, R. Liang, and J. E. Sonier, Physica B  {\bf 326}, 296 (2003).

\bibitem{son97-2} J. E. Sonier, R. F. Kiefl, J. H. Brewer, J. Chakhalian, S. R. Dunsiger, W. A. MacFarlane, R. I. Miller, A. Wong,  G. M. Luke, and J. W. Brill, Phys. Rev. Lett.  {\bf 79}, 1742 (1997).

\bibitem{son04} J. E. Sonier, F. D. Callaghan, R. I. Miller, E. Boaknin, L. Taillefer, R. F. Kiefl, J. H. Brewer, K. F. Poon, and J. D. Brewer, Phys. Rev. Lett.  {\bf 93}, 017002 (2004).

\bibitem{kado} R. Kadono, W. Higemoto, A. Koda, M. Hedo, Y. Inada, Y. Onuki, E. Yamamoto, and Y. Haga,  Phys. Rev. B  {\bf 63}, 224520 (2001).

\bibitem{kadorev} R. Kadono, J. Phys.: Condens. Matter  {\bf 16}, S4421 (2004).

\bibitem{sonson} J. E. Sonier, J. Phys.: Condens. Matter {\bf 16}, S4499 (2004). 

\bibitem{son05} F. D. Callaghan, M. Laulajainen, C. V. Kaiser, and J. E. Sonier, Phys. Rev. Lett.  {\bf 95}, 197001 (2005).

\bibitem{son06} M. Laulajainen, F. D. Callaghan, C. V. Kaiser, and J. E. Sonier, Phys. Rev. B  {\bf 74}, 054511 (2006).

\bibitem{clem75} J. R. Clem, J. Low Temp. Phys. {\bf 18}, 427 (1975). 

\bibitem{hao91} Z. Hao J. R. Clem, M. W. McElfresh, L. Civale, A. P. Malozemoff, and F. Holtzberg, Phys. Rev. B {\bf 43}, 2844 (1991).

\bibitem{brandt95} E. H. Brandt, Rep. Prog  Phys.  {\bf 58}, 1465 (1995).

\bibitem{affl97} I. Affleck, M. Franz, and M. H. S. Amin, Phys. Rev. B {\bf 55}, R704 (1997) .

\bibitem{affl97-2} M. Franz, I. Affleck, and M. H. S. Amin, Phys. Rev. Lett {\bf 79}, 1555 (1997) .

\bibitem{affl98} M. H. S. Amin, I. Affleck, and M. Franz, Phys. Rev. B {\bf 58}, R704 (1998) .

\bibitem{affl00} M. H. S. Amin, M. Franz, and I. Affleck, Phys. Rev. Lett. {\bf 84}, R704 (2000) .

\bibitem{khas05} R. Khasanov, D. G. Eshchenko, D. Di Castro, A. Shengelaya, F. La Mattina, A. Maisuradze, C. Baines, H. Luetkens, J. Karpinski, S. M. Kazakov, and H. Keller, Phys. Rev. B  {\bf 72}, 104504 (2005).

\bibitem{son02} M. D. Lumsden, S. R. Dunsiger, J. E. Sonier, R. I. Miller, R. F. Kiefl, R. Jin, J. He, D. Mandrus, S. T. Bramwell, and J. S. Gardner, Phys. Rev. Lett {\bf 89}, 147002 (2002) .

\bibitem{pros1} D. E. MacLaughlin, J. E. Sonier, R. H. Heffner, O. O. Bernal, B.-L. Young, M. S. Rose, G. D. Morris, E. D. Bauer, T. D. Do, and M. B. Maple, Phys. Rev. Lett. {\bf 89}, 157001 (2002).  L. Shu, D. E. MacLaughlin, R. H. Heffner, F. D. Callaghan, J. E. Sonier, G. D. Morris, O. O. Bernal, A. Bosse, J. E. Anderson, W. M. Yuhasz, N. A. Frederick, and M. B. Maple, Physica B {\bf 374-375}, 247 (2006).

\bibitem{pros3} E. D. Bauer, N. A. Frederick, P.-C. Ho, V. S. Zapf, and M. B. Maple, Phys. Rev. Lett. {\bf 89}, 157001 (2002).

\bibitem{keesom} R. Raderbaugh and P.H. Keesom, Phys. Rev. {\bf 149}, 217 (1966) .

\bibitem{william} S. J. Williamson, Phys. Rev. B {\bf 2}, 3545 (1970) .

\bibitem{sekula} S. T. Sekula and R. H. Kernohan, Phys. Rev. B {\bf 5}, 904 (1972) .

\bibitem{boykin} J. R. Boykin, Jr. and C. J. Bergeron, Jr., Phys. Rev. B {\bf 9}, 2084 (1974) .

\bibitem{hw} E. Helfand and N. R. Werthamer, Phys Rev. {\bf 147}, 288 (1966).

\bibitem{muhl} B. M\"uhlschlegel. Z. Phys.  {\bf 155}, 313 (1959).

\bibitem{YNiHc2} C. Godart, L. C. Gupta, R. Nagarajan, S. K. Dhar, H. Noel, M. Potel, C. Mazumdar, Z. Hossain, C. Levy-Clement, G. Schiffmacher, and B. D. Padalia, R. Vijayaraghavan, Phys. Rev. B {\bf 51}, 489 (1994).

\bibitem{YNiHc2-2}H. Michor, T. Holubar, C. Dusek, and G. Hilscher, Phys. Rev. B {\bf 52}, 16165 (1995).

\bibitem{YNi-n1} K. Izawa, K. Kamata, Y. Nakajima, Y. Matsuda, T. Watanabe, M. Nohara, H. Takagi, P. Thalmeier, and K. Maki, Phys. Rev. Lett. {\bf 89}, 137006 (2002).

\bibitem{YNi-n2} Y. Matsuda and K. Izawa,  Physica C {\bf 388-389}, 487 (2003).

\bibitem{YNi-n3} T. Park, M. B. Salamon, E. M. Choi, H. J. Kim, and S.-I. Lee, Phys. Rev. Lett. {\bf 90}, 177001 (2003).

\bibitem{YNi-n4} K. J. Song, C. Park, S. S. Oh, Y. K. Kwon, J. R. Thompson, D. G. Mandrus, D. McK. Paul, and C. V. Tomy,  Physica C {\bf 398}, 107 (2003).

\bibitem{YNi-n5} T. Watanabe, M. Nohara, T. Hanaguri, and H. Takagi, Phys. Rev. Lett. {\bf 92}, 147002 (2004).

\bibitem{YNi-n6} P. Raychaudhuri, D. Jaiswal-Nagar, G. Sheet, S. Ramakrishnan, and H. Takeya, Phys. Rev. Lett. {\bf 93}, 156802 (2004).

\bibitem{YNi-2g1} T. Terashima, C. Haworth, H. Takeya, S. Uji, H. Aoki, and K. Kadowaki, Phys. Rev. B {\bf 56}, 5120 (1997).

\bibitem{YNi-2g2} S. V. Shulga, S.-L. Drechsler, G. Fuchs, K.-H. MŸller, K. Winzer, M. Heinecke, and K. Krug, Phys. Rev. Lett. {\bf 80}, 1730 (1998).

\bibitem{YNi-2g3} S. Mukhopadhyay, G. Sheet, P. Raychaudhuri, and H. Takeya, Phys. Rev. B {\bf 72}, 014545 (2005).

\bibitem{YNi-2g4} D. L. Bashlakov, Yu. G. Naidyuk, I. K. Yanson, S. C. Wimbush, B. Holzapfel, G. Fuchs, and S.-L. Drechsler, Supercond. Sci. Technol. {\bf 18}, 1094 (2005).

\bibitem{huang} C. L. Huang, J.-Y. Lin, C. P. Sun, T. K. Lee, J. D. Kim, E. M. Choi, S. I. Lee, and H. D. Yang, Phys. Rev. B {\bf 73}, 012502 (2006).

\bibitem{esk} M. R. Eskildsen, P. L. Gammel, B. P. Barber, A. P. Ramirez, D. J. Bishop, N. H. Andersen, K. Mortensen, C. A. Bolle, C. M. Lieber, and P. C. Canfield, Phys. Rev. Lett. {\bf 79}, 487 (1997).

\bibitem{lev} S. J. Levett, C. D. Dewhurst, and D. McK. Paul, Phys. Rev. B {\bf 66}, 014515 (2002).

\bibitem{zhito} M. E. Zhitomirsky and V.-H. Dao, Phys. Rev. B {\bf 69}, 054508 (2004).

\bibitem{mgb1} P. Szab\'{o}, P. Samuely, J. Ka\u{c}mar\u{c}\'{\i}k, T. Klein, J. Marcus, D. Fruchart, S. Miraglia, C. Marcenat, and A. G. M. Jansen, Phys. Rev. B {\bf 87}, 137005 (2001).

\bibitem{mgb2} P. Samuely, P. Szab\'{o}, J. Ka\u{c}mar\u{c}\'{\i}k, T. Klein, and A.G.M. Jansen, Physica C {\bf 385}, 244 (2003).

\end{thebibliography}
\end{document}